\documentclass[aps,prd,twocolumn,tightenlines,preprintnumbers,showpacs,superscriptaddress,notitlepage,multicol,nofootinbib,eqsecnum,floatfix,longbibliography,10pt]{revtex4-1}

\usepackage{amsmath, amssymb,amsfonts}
\usepackage{graphicx}  
\usepackage{dsfont}
\usepackage{enumitem}
\usepackage[pdftex]{color}
\usepackage[utf8]{inputenc}
\usepackage{xspace}
\usepackage{hyperref}
\usepackage{physics}
\usepackage[dvipsnames]{xcolor}
\usepackage{xcolor,soul}
\usepackage{xfrac}
\usepackage{ulem}
\usepackage{soul}
\hypersetup{
    colorlinks=true,  
    linkcolor=blue,   
    citecolor=blue,   
    filecolor=blue,   
    urlcolor=blue     
}

\newcommand\cern{CERN, Theoretical Physics Department, 1211 Geneva 23, Switzerland}
\newcommand{\dias}{
    Danish Institute for Advanced Study, 
    University of Southern Denmark,  5230 Odense M, Denmark
}
\newcommand{\cpthreemath}{
    Quantum Theory Center ($\hbar$QTC), Department of Mathematics and Computer Science,
    University of Southern Denmark, 
    5230 Odense M, Denmark
}
\newcommand{\avg}[1]{\left< #1 \right>}

\usepackage{pifont}

\begin{document}

\author{Pietro~Butti}\affiliation{\cpthreemath}\affiliation{\dias}
\author{Michele~Della~Morte}\affiliation{\cpthreemath}

\author{Benjamin~J{\"ager}}\affiliation{\cpthreemath}\affiliation{\dias}
\author{Sofie~Martins}\affiliation{\cpthreemath}
\author{J.~Tobias~Tsang}\affiliation{\cern}

\date{\today}

\title{Comparison of smoothening flows for the topological charge in QCD-like theories}

\begin{abstract}
  We investigate properties of the topological charge for several $SU(N_C)$
  gauge field ensembles for $N_C=4,5,6$ with a single fermion in the two-index
  anti-symmetric representation, covering multiple lattice spacings at otherwise
  approximately constant physical parameters. Comparing the topological charge
  defined by the Wilson flow and the over-improved DBW2 flow we find that
  already at small flow times the latter stabilises on discrete values.  We
  provide evidence that as the lattice spacing is lowered the Wilson flow also
  separates into discrete sectors at earlier flow times. Adopting the DBW2
  definition in the remainder of the analysis, we do not see any evidence of
  fractional topological charges, which could in principle appear at finite
  lattice spacing.
\end{abstract}

\maketitle
\preprint{CERN-TH-2025-079}

\section{Introduction and Motivations}

We recently started exploring the spectrum of $SU(N_C)$ {\it orientifold}
theories with one dynamical quark flavour in the two-index (anti-)symmetric
representation~\cite{DellaMorte:2023ylq, DellaMorte:2025tks, Martins:2023kcj,
  Jaeger:2022ypq, Ziegler:2021nbl}.  In the large $N_C$ limit at fixed 't Hooft
coupling, i.e. in the Corrigan-Ramond limit~\cite{Corrigan:1979xf}, the theory
should approach Super-Yang-Mills as first conjectured in
Refs.~\cite{Armoni:2003gp,Armoni:2003fb}. Predictions for the ratio of masses of
the lightest scalar and pseudoscalar mesons have been obtained using either
effective field theory or string theory duality
approaches~\cite{Sannino:2003xe,Armoni:2005qr} and such predictions can be
compared to lattice results, as done in Ref.~\cite{Sannino:2024xwj} using our
result from Ref.~\cite{DellaMorte:2023ylq}.

For $N_C=3$ the two-index anti-symmetric representation coincides with the
fundamental representation and hence standard codes optimised for QCD can be
used. For $N_C \neq 3$ this is no longer the case, and with increasing $N_C$
simulations become more costly.  Naively, the cost of a matrix-vector
multiplication grows quadratically with $N_C$, while for matrix-matrix
multiplications the scaling is cubic. One therefore expects an overall scaling
with $N_C$ slightly faster than quadratic, assuming that matrix-vector
multiplications dominate the cost. This is consistent with our findings in
Ref.~\cite{Martins:2023kcj}.  In addition, we are entering a largely unexplored
territory concerning the choices of bare (Lagrangian) as well as algorithmic
parameters. To the best of our knowledge no other study exists where lattice
simulations of the exact same theories considered here are discussed. One of our
main concerns, given the cost of the simulations, is critical slowing down,
known to become more severe as $N_C$ is increased~\cite{DelDebbio:2002xa}. This
is usually monitored by looking at the Monte-Carlo distribution of a slow
(i.e. with long autocorrelation) observable, typically the topological charge.

The first goal of this paper is to select a numerical definition of the
topological charge which is cheap and stable, in a sense that will be clarified
below.  Secondly, the study at hand is part of our wider programme to determine
the mass spectrum and properties of orientifold theories in the large $N_C$
limit.  Amongst such properties, here we focus on the behaviour of the
topological charge for different choices of $N_C$ and investigate the existence
of fractional values.  Indeed, for the two-index anti-symmetric representation,
the winding number (i.e. the topological charge) of a gauge configuration is in
general expected to be quantised in units of $1/(N_C-2)$, see also
Ref.~\cite{Leutwyler:1992yt} for a discussion of a single flavour in the adjoint
representation.  However, with periodic boundary conditions (as used here) and
for sufficiently smooth configurations the topological charge should take
integer values, as first shown in Ref.~\cite{Luscher:1981zq}. When different
boundary conditions, such as 't Hooft twisted boundary conditions, are used
fractional charges do instead survive in the continuum
limit~\cite{tHooft:1979rtg,GarciaPerez:2000aiw,Gonzalez-Arroyo:2019wpu}.  In our
case we hence expect non-integer charges to appear as cut-off effects, if at
all.  A similar study was conducted in Ref.~\cite{Fodor:2009nh} for the SU(2)
case of a single flavour in the two-index symmetric representation (sextet
model). The study was quenched, with the topological charge defined through the
index-theorem~\cite{Atiyah:1968mp} and the conclusion was indeed that
configurations with fractional charge can occur, but disappear as the continuum
limit is approached.

Here we determine the topological charge using a gluonic definition from
smoothened (gradient-flowed) gauge fields.  We systematically study the
dependence of the charge on the amount of smoothening applied, as well as on the
discretisation scheme employed for the flow equations.
Ref.~\cite{Tanizaki:2024zsu} performs a similar study but at a single lattice
spacing in pure Yang-Mills for $SU(2)$. The findings in this work extend beyond
that by varying the lattice spacing, including dynamical fermions and exploring
different gauge groups.

For $N_C=4$ and $N_C=5$ we generated a number of ensembles with approximately
tuned spatial extents and connected pseudoscalar ($M_\pi^\mathrm{conn.}$)
masses, but varying lattice spacings. For $N_C=6$ we generated a single ensemble
with similar volume and mass and an intermediate lattice spacing. The generation
and properties of these ensembles and the smoothening flows are described in
more detail in Section~\ref{sec:comp}. In Section~\ref{sec:anal} we determine
the lattice spacings and investigate the topological charge and susceptibility
as a function of flow time and the choice of flow. Finally, in
Section~\ref{sec:conc} we summarise our findings and provide an outlook.

\section{Computational setup\label{sec:comp}}

\subsection{Ensemble Parameters 
\label{subsec:ens}}

The gauge configurations were generated using the \texttt{HiRep} code
package~\cite{DelDebbio:2008zf,Martins:2024dew,Martins:2024sdd,Drach:2025eyg}. For
the gluonic part of the action, we employ the tree-level improved
L\"uscher-Weisz (LW) gauge action~\cite{Luscher:1985zq}. The fermionic part
contains a single Dirac flavour in the two-index anti-symmetric
representation. We use the Wilson formulation with a tree-level improved clover
term, i.e. setting $c_{SW} =1$. Since our setup contains a single Dirac flavour,
the Rational Hybrid Monte-Carlo (RHMC) algorithm~\cite{Kennedy:2012gk} is
required. For the numerical integration of the RHMC trajectories, we adopt a
4th-order Omelyan integrator~\cite{Omelyan:2002qkh} with a trajectory length of
$\tau = 2$\, molecular dynamic units (MDUs). The number of integration steps in
the fermionic force (20 - 26 for $N_C = 4$, 30 - 36 for $N_C=5$, and 36 for $N_C
=6$) is tuned to ensure a high acceptance rate. In the gauge sector we also use
a 4th-order Omelyan integrator with two steps per fermion update. To reduce
autocorrelation, we save every fourth configuration, which results in a
separation of $8$\,MDUs between consecutively saved gauge configurations. For
thermalisation, the first 200 trajectories are discarded. All gauge
configurations were generated on GPUs (AMD MI250x) using the LUMI-G partition,
while all measurements were carried out on CPUs. Table~\ref{tab:ens} provides an
overview of the ensembles generated for this study. The relevant input files are
provided in the arXiv submission.

\begin{table*}
\caption{Parameters of the ensembles used in this study. The temporal extent is
  fixed to be $T = 3 L$ for all ensembles. Quoted uncertainties are statistical
  only.}
\begin{tabular}{l|cccccccccc}
     name & $N_C$ & $\beta$ & $\kappa$ & $L/a$ & $t_0/a^2$ & $aM_\pi^\mathrm{conn.}$ & $L/\sqrt{t_0}$ & $M_\pi^\mathrm{conn.} \sqrt{t_0}$ & $N_\mathrm{conf}$
     \\\hline\hline
     N4L12  & 4&  7.1 & 0.15770 & 12 & 0.8691(12) &  0.6481(6) &12.8720(90)& 0.6042(7)  & 648\\
     N4L14  & 4&  7.2 & 0.15651 & 14 & 1.3641(17) & 0.5101(5) &11.9868(75)& 0.5958(7) & 677\\
     N4L16  & 4&  7.3 & 0.15525 & 16 & 1.8592(19) & 0.4350(4) &11.7344(60)& 0.5931(6) & 640\\
     N4L18  & 4&  7.4 & 0.15401 & 18 & 2.3885(21) & 0.3949(3) &11.6470(51)& 0.6103(5) & 583\\  
     \hline
     N5L12 & 5& 11.3 & 0.16089 & 12 & 1.1171(11) & 0.7571(3) &11.3536(55)& 0.8002(5) & 1081\\
     N5L14 & 5& 11.4 & 0.16026 & 14 & 1.4800(14) & 0.6624(3) &11.5078(53)& 0.8058(5) & 704\\
     N5L16a & 5& 11.5 & 0.15959 & 16 & 1.8422(15) & 0.5581(3) &11.7884(49)& 0.7575(5) & 565\\
     N5L16b &5& 11.55& 0.15929 & 16 & 2.0409(21) & 0.5922(2) &11.1997(58)& 0.8460(5) & 469\\\hline
     N6L16 & 6& 16.5 & 0.16200 & 16 & 1.33944(64)& 0.8354(1) &13.8248(33)& 0.9668(3) & 742\\
\end{tabular}
\label{tab:ens}
\end{table*}

Since the simulated masses are rather heavy (see Table~\ref{tab:ens}), based on
our experience in Ref.~\cite{DellaMorte:2023ylq}, we do not expect these
simulations to be affected by a sign problem. In the future we will supplement
the ensembles in this work with ensembles at lighter-quark masses and perform
extrapolations of observables to the massless limit. In this context we will
perform a study of the severity of the sign problem for the lightest simulated
quark masses, analogous to that of Ref.~\cite{DellaMorte:2023ylq}.

\subsection{Flow definitions \label{subsec:defs}}
The gradient flow is a standard tool in Lattice Field Theory, consisting in
evolving gauge fields in a fictitious 5th dimension typically referred to as
\textit{flow time} $t$. This evolution is governed by a gauge-covariant
diffusion equation
\begin{equation}\label{flow}
    \dv{B_{\mu,x}(t)}{t} = D_\nu G_{\mu\nu,x}(t)\,,
\end{equation}
with initial conditions $B_{\mu,x}(t=0)=A_\mu(x)$, $G_{\mu\nu,x}(t)$ being the
field strength of the field $B_{\mu,x}(t)$ at flow time $t$. The core idea is
that, as shown in the original paper~\cite{Luscher:2010iy}, the flow leads to an
effective smearing of the gauge configurations over a length scale $\sqrt{8t}$,
suppressing UV fluctuations and automatically rendering a large class of
composite operators renormalised at positive flow times.

On the lattice, different discretisations of Eq.~\eqref{flow} are possible,
which can be re-written as
\begin{equation}
    a^2\qty(\frac{d}{d t} U_{x,\mu}(t)) U_{\mu,x}(t)^\dagger = -g_0^2 \partial_{\mu,x}  S_\mathrm{flow}(U),
\end{equation}
where $\partial_{\mu,x}$ is the Lie-algebra valued derivative with respect to
$U_{\mu,x}(t)$. $S_\text{flow}$, also referred to as \textit{kernel action},
represents some discretised version of the gauge action that is used to
integrate the flow equations and which, in principle, can be different from the
one used to generate the configurations. One possible choice is to consider
$\order{a^2}$-improved actions defined with an appropriate combination of 6-link
Wilson loops, beside the usual 4-link (plaquette)
loop~\cite{Luscher:1984xn}. Considering only 4 and 6-links Wilson loops lying on
a plane, we can define the kernel action as
\begin{equation}\label{Sc0c1}
    S_\mathrm{flow} = c_0 S_\mathrm{plaquette} + c_1 S_\mathrm{rectangle}\,,
\end{equation}
parameterised by 2 coefficients $c_0$ and $c_1$ constrained to satisfy
\begin{equation}
    c_0 + 8c_1 = 1
\end{equation}
in order to ensure that in the classical continuum limit, $S_\text{flow}$
recovers the standard Yang-Mills action.  Among others, popular choices include:
$c_1=0$ (the standard plaquette action, commonly referred to as \textit{Wilson}
flow), $c_1=-0.311$ (also called Iwasaki~\cite{Itoh:1984ym}), $c_1=-1/12$
(tree-level improved L\"uscher-Weisz (LW)~\cite{Luscher:1985zq}) and
$c_1=-1.4088$ (DBW2~\cite{QCD-TARO:1999mox,QCD-TARO:1996lyt}).  In this work, we
will consider the case of Wilson and DBW2 flows. A study of all the four
mentioned choices for $c_1$ in pure $SU(2)$ Yang-Mills appeared recently in
Ref.~\cite{Tanizaki:2024zsu}.  In the same reference, the action for a
one-instanton configuration is tree-level evaluated to $\order{a^2}$. The
relative correction to the continuum value is given by $-(1+12c_1)/5 \times
\left(a/\rho\right)^2$, $\rho$ being the radius of the instanton.  As a result,
for $1+12c_1>0$ the flow-action is expected to favour (by lattice artefacts)
small size instantons, whereas for $1+12c_1<0$ larger instantons should be
preferred.  The latter choice leads to $\it {over-improvement}$ as first
discussed along the lines above but in the framework of cooling in
Ref.~\cite{GarciaPerez:1993lic}.  The combination $1+12c_1$ in particular
corresponds to the coefficient $\epsilon$ in Ref.~\cite{GarciaPerez:1993lic}
when only plaquettes and rectangles are considered.

We modified the \texttt{HiRep} code to include the rectangle terms in the kernel
action.  All flow measurements were performed using a third-order Runge-Kutta
integration scheme with a fixed step size of $0.01$ in lattice units.  We tested
our \texttt{HiRep}-based implementation of the DBW2 flow by comparing our
results to data obtained using the implementation in
\texttt{LatticeGPU}~\cite{Catumba:2025jae} on a test configuration. The results
agree to machine precision.

As mentioned above, the length scale over which the configurations are smeared
is $\sqrt{8t}$, also for the case of over-improved actions, as discussed in
Appendix B of Ref.~\cite{Tanizaki:2024zsu}. The flow time $t$ should therefore
be chosen to satisfy
\begin{equation}
    2a \ll \sqrt{8t} \ll \frac{L}{2}\,,
    \label{eq:bounds}
\end{equation}
in order to sufficiently suppress discretisation (left inequality) and finite
volume (right inequality) effects.  As a remark, in the initial stages of this
study we also considered the Iwasaki-flow which displays qualitatively similar
features to DBW2. However, for the volumes at hand we found that on a subset of
configurations the topological charge reached a stable value only for $t$
uncomfortably close to the right bound in the equation above.

\begin{figure*}
  \includegraphics[width=\textwidth]{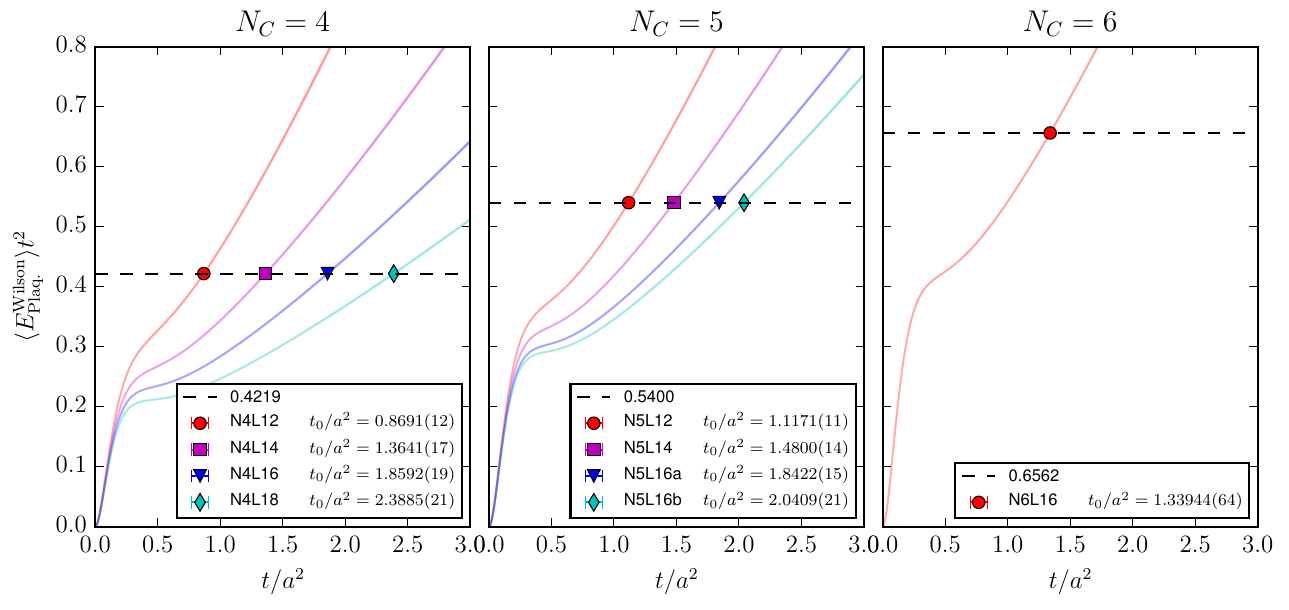}
  \caption{Determination of $t_0/a^2$ on all ensembles.}
  \label{fig:scale}
\end{figure*}
\section{Analysis \label{sec:anal}}
\subsection{Scale setting \label{subsec:scale}}
To set the scale of our simulations, we integrate the Wilson flow and compute
the expectation value of the flowed energy density $\expval{E_x(t)}=\expval{\tr
  G_{\mu\nu,x}(t)G_{\mu\nu,x}(t)}$, choosing the plaquette discretisation
$E_\text{plaq}$ for the action density. The relative lattice scale is usually
set by determining $t_0$ for which $\avg{E_\mathrm{plaq}} t_0^2$ takes a
particular value. For the case of $N_C=3$ this value is typically chosen to be
$0.3$. Generalising this for $N_C>3$ whilst accounting for the leading $N_C$
dependence of $\avg{E}$, and hence staying a fixed 't Hooft coupling, requires a
rescaling by $(N_C^2-1)/N_C$ and so yields
\begin{equation}
  \avg{E_\mathrm{plaq}} t_0^2 = 0.3 \left(\frac{N_C^2-1}{N_C}\right)\frac{3}{8}\,.
\end{equation}
This value can then be related to an absolute scale via the determination of a
reference $t^\mathrm{ref}_0$, obtained from a scale-setting fit. Since (to the
best of our knowledge) no such determination exists for a single flavour
($N_f=1$), for concreteness we use the average of the
$N_f=0$~\cite{Luscher:2010iy} and $N_f=2$~\cite{Bruno:2013gha} results in QCD
obtaining
\begin{equation}
  \sqrt{8t_0^\mathrm{ref}} = 0.45\,\mathrm{fm}\,.
\end{equation}
This yields lattice spacings $a\in [0.10,0.17]\,\mathrm{fm}$
($[0.11,0.15]\,\mathrm{fm}$) for the $N_C=4$ ($N_C=5$) ensembles and $a \sim
0.14\,\mathrm{fm}$ for $N_C=6$. We stress that the absolute scale is not
essential for this work, but useful to have an approximate estimate of the
lattice spacings, masses and volumes under consideration. The values listed in
Table~\ref{tab:ens} contain statistical uncertainties only which are estimated
from a bootstrap analysis, not taking autocorrelations into account. Applying
the $\Gamma$-method~\cite{Wolff:2003sm} instead leads to somewhat larger
uncertainties, possibly indicating the presence of autocorrelations in the
data. In the worst case (on the \texttt{N6L16} ensemble) the uncertainty
estimate from the $\Gamma$-method is a factor 2.5 larger. Since in this work we
do not present precision observables this is not expected to have any impact at
this stage, but we will carefully investigate this in future studies.

\subsection{Observables \label{subsec:obs}}
For this work, the main observable of interest is the topological charge $Q$,
defined in the classical continuum limit as the integral of the topological
charge density $q(x)$
\begin{equation}\label{topq_cont}
  Q = \int\dd[4]{x} q(x)\,,
\end{equation}
where $q$ is defined as
\begin{equation}\label{topqd}
  q(x) = \frac{1}{32\pi^2}\epsilon_{\mu\nu\rho\sigma}\tr\qty[F_{\mu\nu}F_{\rho\sigma}]\,,
\end{equation}
and $F_{\mu\nu}$ is the field strength tensor of the unflowed field.  On the
lattice, several discretisations of the topological charge density can be taken,
all leading to Eq.~\eqref{topqd} in the classical $a\rightarrow 0$ limit. One
possible definition is\footnote{For simplicity, we adopt the symbol $Q$ to
  indicate both the continuum topological charge in Eq.~\eqref{topq_cont} and
  its lattice discretised version in Eq.~\eqref{topq}.}
\begin{equation}\label{topq}
  Q = \frac{1}{32 \pi^2} \sum_x \epsilon_{\mu \nu \rho \sigma} \tr \qty[ \hat C_{\mu \nu}(x) \hat C_{\rho \sigma}(x) ]\,,
\end{equation}
where for concreteness $\hat C$ is the clover discretisation of the field
strength which is used in our work.  In principle, regardless of the
discretisation, the lattice topological charge must be properly
renormalised. Alternatively, some smoothening techniques of the gauge
configurations can be employed. Among others, popular choices include
cooling~\cite{Berg:1981nw, Iwasaki:1983bv, Itoh:1984pr, Teper:1985rb,
  Campostrini:1989dh, Alles:2000sc}, stout-smearing~\cite{Morningstar:2003gk}
and the gradient flow~\cite{Luscher:2010iy, Ce:2015qha}.\footnote{Detailed
  comparisons between these methods can be found, e.g. in
  Refs.~\cite{Alles:2000sc, Bonati:2014tqa, Alexandrou:2015yba,
    Alexandrou:2017hqw}.}  In practice, one also has to control the amount of
smoothening applied to the gauge configurations. In the case of the gradient
flow one needs to choose the flow time $t_Q$ at which $Q$ is computed. Whilst
Ref.~\cite{Luscher:2010iy} suggested to use $t_Q\simeq t_0$, we will explore the
impact of choosing different kernel actions to integrate the flow as in
Eq.~\eqref{Sc0c1} as well as the influence of $t_Q$. On the lattice, even after
several smoothening steps are applied $Q$ is not expected to take discrete
values but rather to be distributed over real numbers clustered around
peaks.\footnote{This can be corrected for using rounded definitions, see
  e.g. Refs.~\cite{DelDebbio:2002xa,Bonati:2015sqt}.}
  
The cumulants of the distribution of $Q$ can be used to compute physical
quantities, such as the \textit{topological susceptibility} $\chi$, defined as
\begin{equation}\label{chi}
  \chi = \lim_{a\rightarrow 0} \lim_{V_4\rightarrow\infty} \frac{\expval{Q^2}}{V_4}\,,
\end{equation}
where $V_4 = L^3T$ is the 4-volume in physical units.  Another observable that
we will consider is the so-called \textit{smoothness parameter} defined in
Refs.~\cite{Luscher:1981zq,Luscher:2010iy} and adapted to our case as
\begin{equation}
  h(p) = \mathrm{Re}\, \mathrm{Tr} \left( \mathds{1} - \prod_{(x,\mu)\in p}U_{x,\mu} (t)\right)\,,
  \label{eq:h}
\end{equation}
$p$ being a given plaquette. We distinguish between 
\begin{equation}
  \begin{aligned}
    h_\mathrm{max} &= \underset{p}{\mathrm{max}}\,(h)\,,    \\
    h_\mathrm{avg} &= \underset{p}{\mathrm{avg}}\,(h)\,.
    \label{eq:hdefs}
  \end{aligned}
\end{equation}
Since the flow is a smoothening procedure, $h_\mathrm{avg}$ should monotonically
decrease as a function of flow time. However, as we will see, $h_\mathrm{max}$
is a local object and doesn't necessarily have a monotonic behaviour.  In order
for the topological sector of gauge configurations to be well-defined and
characterisable, the smoothness parameter $h_\text{max}$ has to be smaller than
a critical value of order $10^{-1}$ for the theories considered
here~\cite{Luscher:1981zq,Luscher:2010iy}. Topological sectors cannot be
connected by continuous deformations, and $h_\text{max}$ must exceed such a
critical value when gauge configurations move among sectors.  However, as
discussed in Ref.~\cite{Luscher:2010iy}, the occurrence of such configurations
decreases proportionally to $a^{6}$, which leads to the problem of topological
freezing as the lattice spacing is reduced. This makes clear that studies of
topological properties of a lattice gauge theory suffer a window problem. On the
one side the lattice spacing should be fine enough such that configurations are
sufficiently smooth to be classified according to a lattice version of the
topological charge, while on the other side topological freezing should be
avoided in order to have an adequate sampling of all the sectors.

\subsection{Determination of the topological charge}
\begin{figure}
  \centering
  \includegraphics[width=\columnwidth]{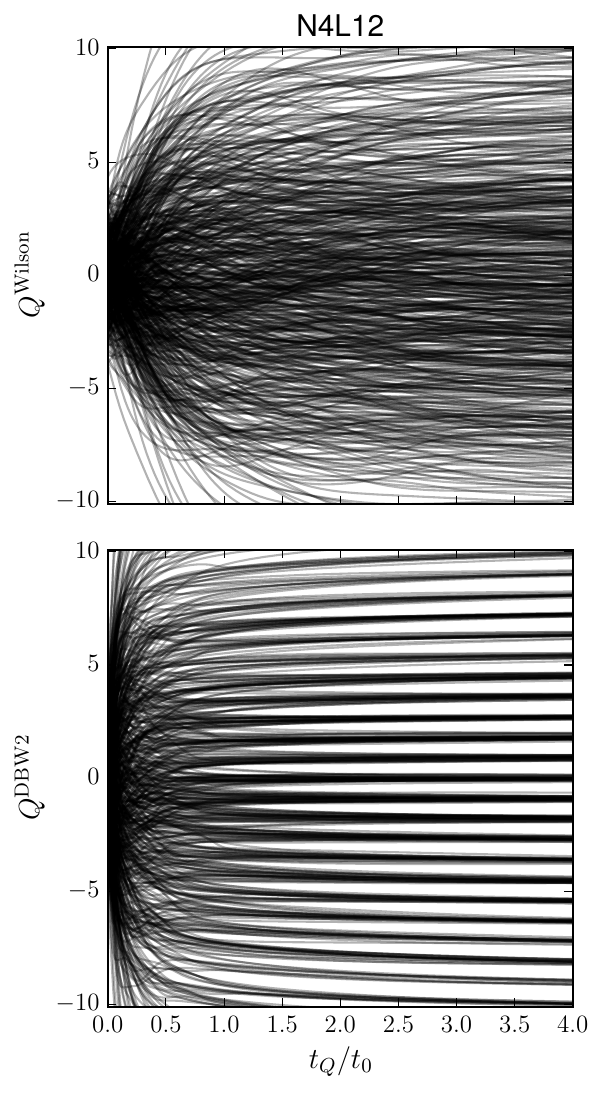}
  \caption{Topological charge determined using the Wilson flow (top) and the
    DBW2 flow (bottom) on the coarsest $N_C=4$ ensemble.}
  \label{fig:compQN4L12}
\end{figure}

\begin{figure}
  \centering
  \includegraphics[width=\columnwidth]{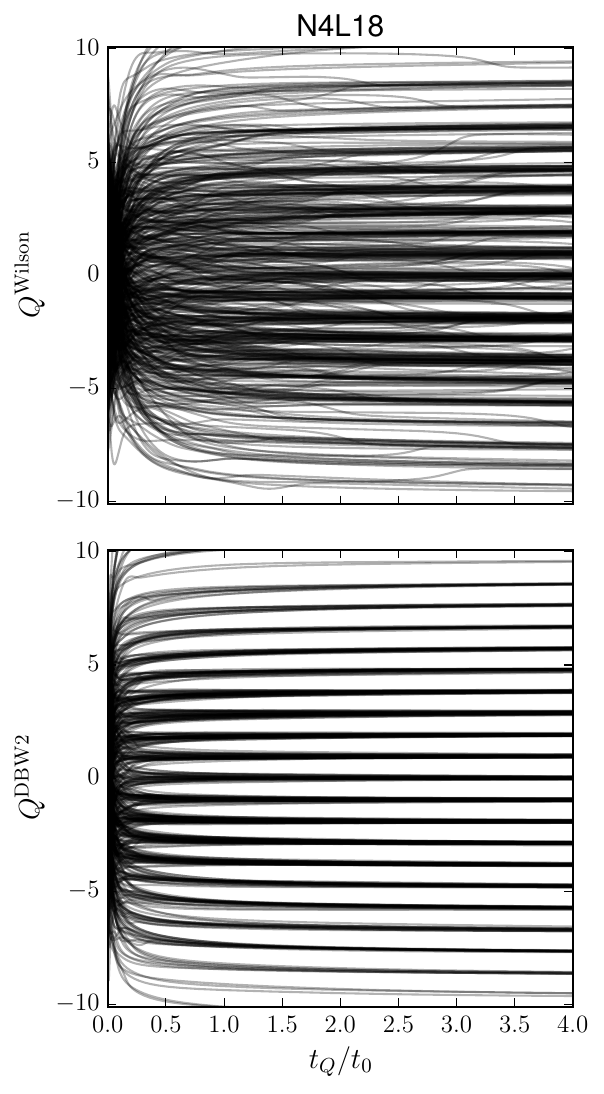}
  \caption{Topological charge determined using the Wilson flow (top) and the
    DBW2 flow (bottom) on the finest $N_C=4$ ensemble.}
  \label{fig:compQN4L18}
\end{figure}

At each positive flow time $t$ we compute the topological charge for the
Wilson-flowed and the DBW2-flowed configurations. Since we want to investigate
the existence (or absence) of fractional topological charges, on a given
configuration we desire a definition of $Q$ which admits a window in flow time
for which $Q$ is insensitive of the exact choice of $t_Q$ whilst adhering to the
bounds in Eq.~\eqref{eq:bounds}.

The top (bottom) panel of Fig.~\ref{fig:compQN4L12} shows the topological charge
as defined by the Wilson (DBW2) flow as a function of flow time for our coarsest
ensemble (\texttt{N4L12}). Each line corresponds to a different
configuration. We observe that for this ensemble the Wilson flow does not admit
such a definition, and instead find that even for large $t_Q$ the topological
charge fluctuates as a function of flow time. In contrast to this, the DBW2 flow
does admit such an assignment.  When instead considering our finest ensemble
(\texttt{N4L18}), as shown in Fig.~\ref{fig:compQN4L18}, we note that the rate
of jumps between topological sectors decreases for both flows. For the Wilson
flow, several configurations maintain an asymptotically stable value, but some
jumps can still be seen even for the largest plotted flow times. For the DBW2
flow, the configurations quickly stabilise into discrete topological sectors.

\begin{figure}
\centering
\includegraphics[width=\columnwidth]{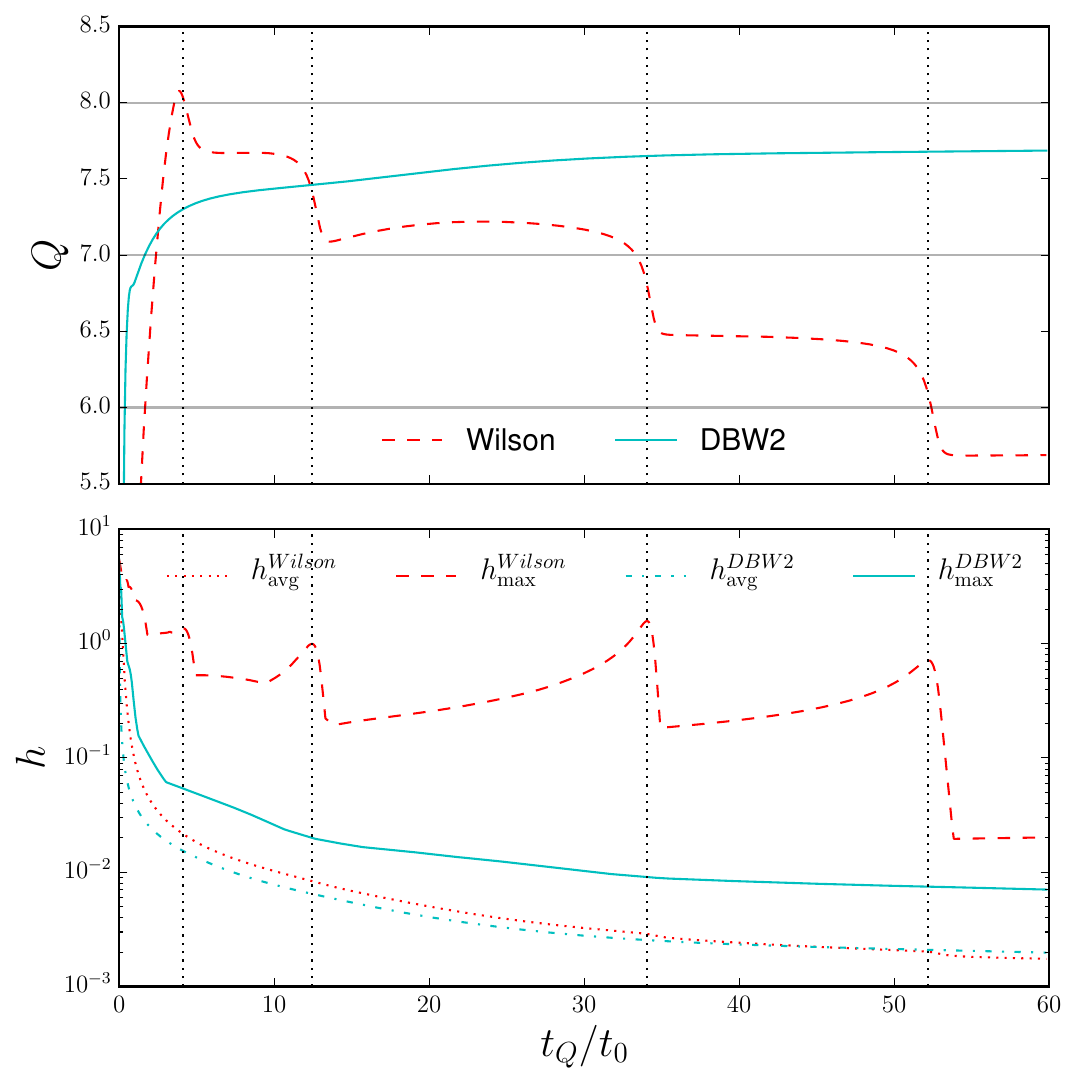}
\caption{Topological charge (top) as well as $h_\mathrm{avg}$ and
  $h_\mathrm{max}$ (bottom) for the Wilson flow (red, dashed and dotted) and the
  DBW2 flow (cyan, solid and dash-dotted) as a function of very long flow times
  and on a single configuration of the \texttt{N4L12} ensemble. The vertical
  dotted lines correspond to the location of local maxima in
  $h_\mathrm{max}^\mathrm{Wilson}$.}
\label{fig:topo_hmax}
\end{figure}
When rewriting the upper bound on $t_Q$ in Eq.~\eqref{eq:bounds} above in terms
of $t_Q/t_0$ we find
\begin{equation}
    \frac{t_Q}{t_0} \ll \frac{1}{32}\frac{L^2}{t_0}\,.\label{eq:boundstt0}
\end{equation}
Substituting the corresponding numbers from Table~\ref{tab:ens} we find the
expected bounds to be $t_Q/t_0 \ll 5.18$ for \texttt{N4L12} and $t_Q/t_0 \ll
4.24$ for \texttt{N4L18}.  Whilst we do not formally quantify this, by visual
inspection of our data, e.g. in the top panel of Fig.~\ref{fig:compQN4L18}, it
appears that for the Wilson flow the number of jumps between Q-values on any
given configuration decreases when comparing the interval $t_0 <t< 2t_0$ to
$3t_0<t<4t_0$.  Indeed for the latter interval a clear clustering around
discrete values is visible, which is not the case for the first. In the
following we explore the behaviour as the flow time is increased further towards
and beyond the upper bound in Eq.~\eqref{eq:boundstt0}. Whilst we are aware that
this eventually introduces non-localities, we are curious about the qualitative
features this induces.  Figure~\ref{fig:topo_hmax} shows the topological charge
$Q$, $h_\mathrm{max}$ and $h_{\mathrm{avg}}$ as defined in Eq.~\eqref{eq:hdefs}
as a function of $t_Q/t_0$ for very long flow times on a representative
configuration of the \texttt{N4L12} ensemble. As expected for a smoothening
flow, the value of $h_\mathrm{avg}$ monotonically decreases as a function of
flow time for both choices of flow. For the DBW2 flow, this also holds for
$h_\mathrm{max}$, indicating an approximately uniform smoothening, whilst for
the Wilson flow this is not the case. Instead there are several occurrences
where $h_\mathrm{max}^\mathrm{Wilson}$ suddenly increases before dropping again
(note the logarithmic scale in the bottom panel). Since the average of $h$ still
decreases, this must be a local (at the lattice scale) effect. This observation
is consistent with the expectation for non-over-improved actions, as they do not
suppress small instantons, as discussed in Sec.~\ref{subsec:defs} and
Ref.~\cite{GarciaPerez:1993lic}.  Turning our attention to the topological
charge we find that $Q^\mathrm{DBW2}$ quickly becomes largely independent of
flow time and slowly approaches an asymptotic value. We will discuss the fact
that this value does not appear to be an integer in due course. The topological
charge $Q^\mathrm{Wilson}$ instead displays meta-stable plateaus over some time
scales, but continues to jump between different topological sectors even for
very late times. In order to assess the existence of fractional topological
charges, we hence prefer the DBW2 flow. From a closer look at the spikes in
$h_\mathrm{max}^\mathrm{Wilson}$, we find that they correlate with jumps in
$Q^\mathrm{Wilson}(t_Q)$ (as indicated by the vertical lines).

\begin{figure*}
\centering
\includegraphics[width=\columnwidth]{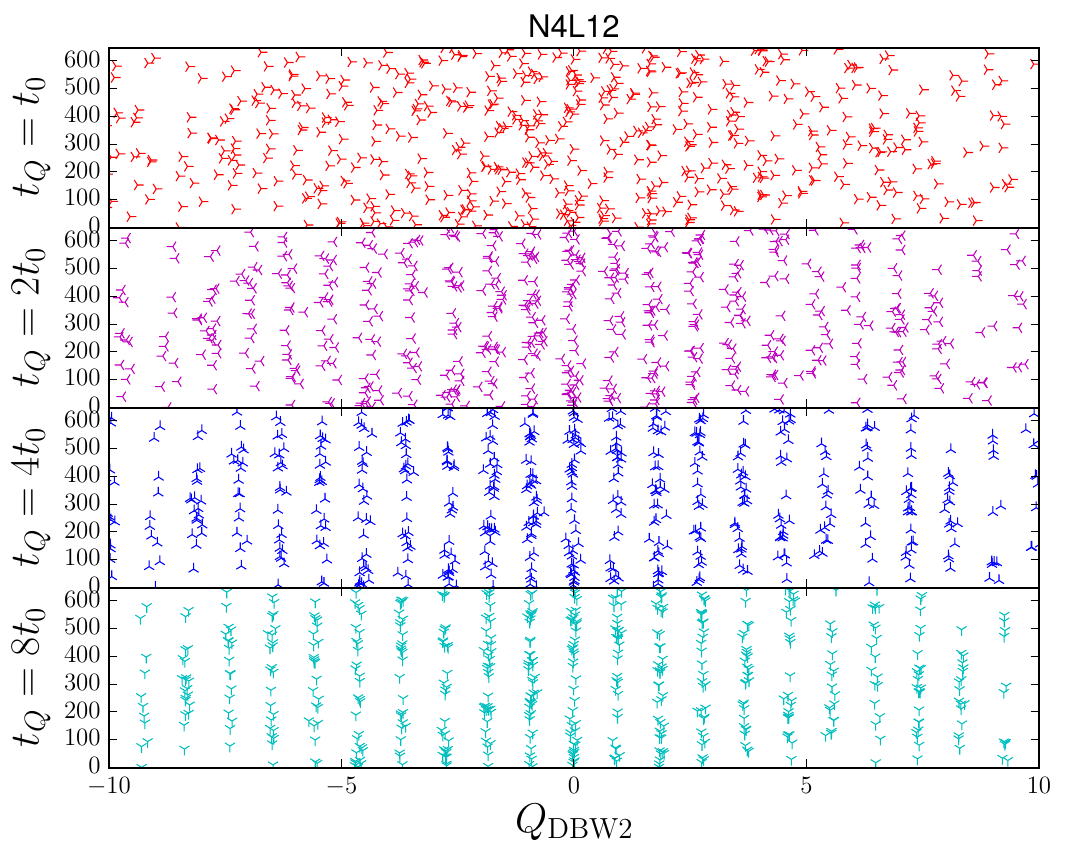}
\includegraphics[width=\columnwidth]{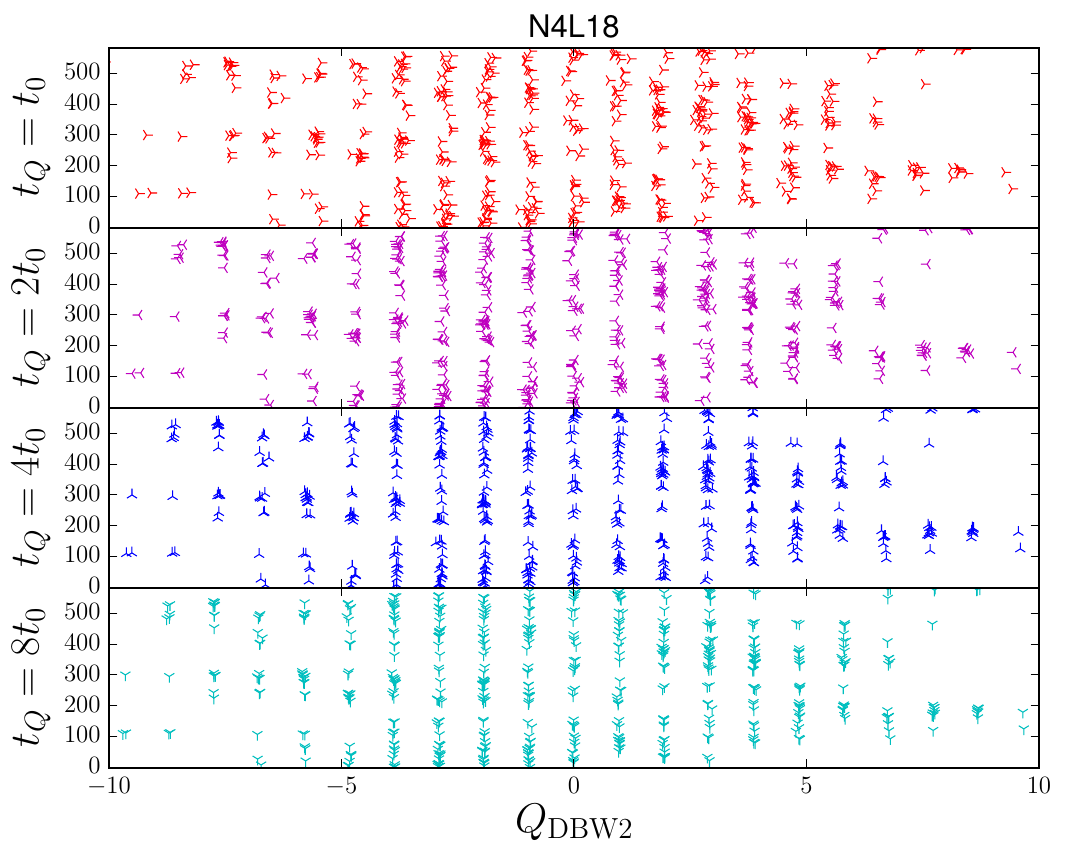}
\caption{Topological charge determination with the DBW2 flow for the
  \texttt{N4L12} (left) and \texttt{N4L18} (right) ensembles at 4 flow times
  (from top to bottom $t_Q=t_0$, $2t_0$, $4t_0$ and $8t_0$) plotted
  configuration by configuration.}
\label{fig:topo_by_conf}
\end{figure*}
In Figure~\ref{fig:topo_by_conf} we depict the topological charge determined by
the DBW2 flow configuration-by-configuration for flow times $t_Q=t_0$, $2t_0$,
$4t_0$ and $8t_0$. As discussed before, we notice that discrete topological
sectors form, but that they do not exactly coincide with integer values. Instead
they appear to be multiples of a number close to, but slightly smaller than,
unity. We observed in Figures~\ref{fig:compQN4L12} and \ref{fig:compQN4L18},
that (within a given discrete sector) the topological charge continues to slowly
grow in magnitude as a function of flow time. In Figure~\ref{fig:topo_hmax} we
saw that this trend continues even for very large flow times but still not
reaching integer values.  When comparing the \texttt{N4L12} and the
\texttt{N4L18} ensembles in Figure~\ref{fig:topo_by_conf}, we notice that with
decreasing lattice spacing, the topological sectors approach integer
values. Both of these effects contribute to the unit of discretisation being
smaller than one. In order to faithfully capture the discrete sectors when
representing the data in histograms, in the following we use bin widths that
reproduce this feature. In particular we will define the bin width $\epsilon$
such that $2\epsilon$ to a good approximation corresponds to the separation
between adjacent topological sectors.

\begin{figure*}
\centering
\includegraphics[width=.32\textwidth]{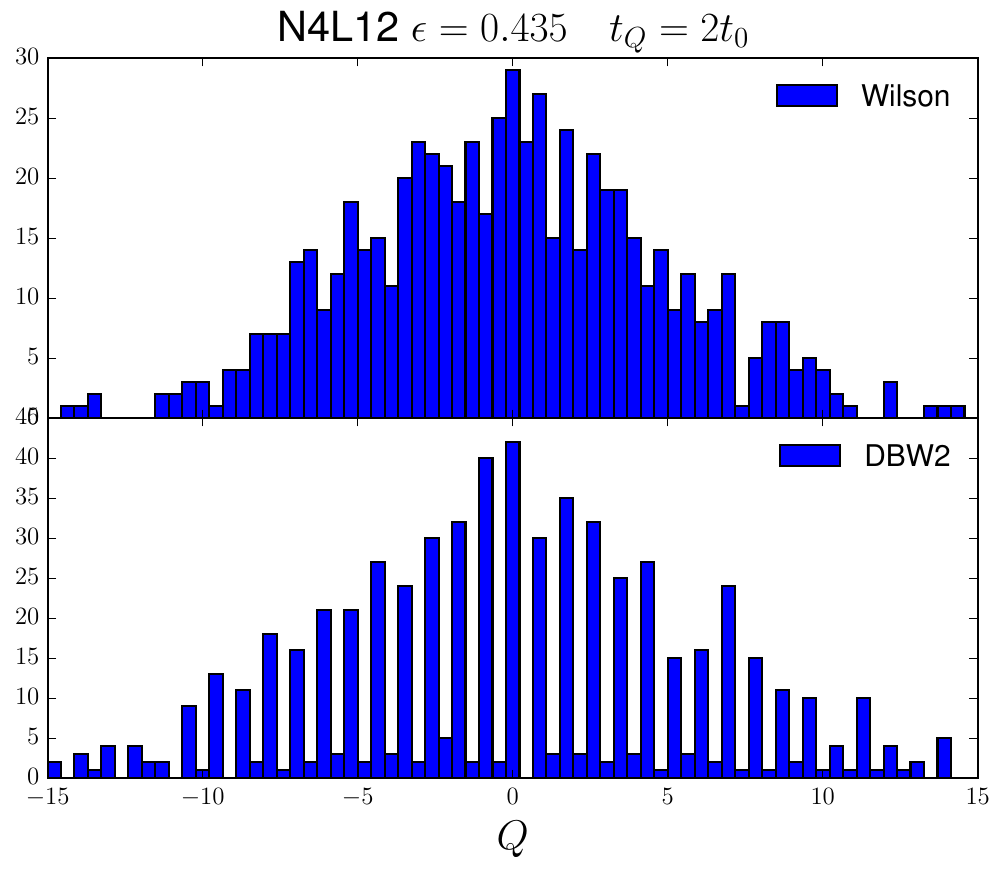}
\includegraphics[width=.32\textwidth]{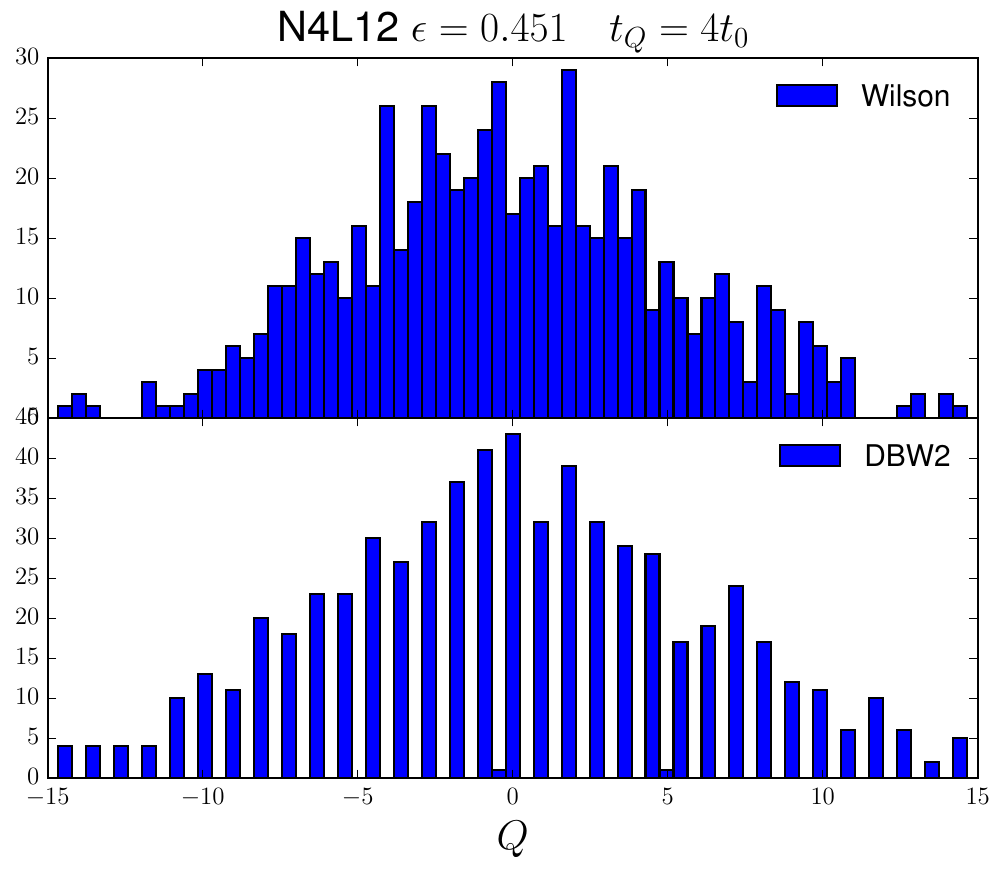}
\includegraphics[width=.32\textwidth]{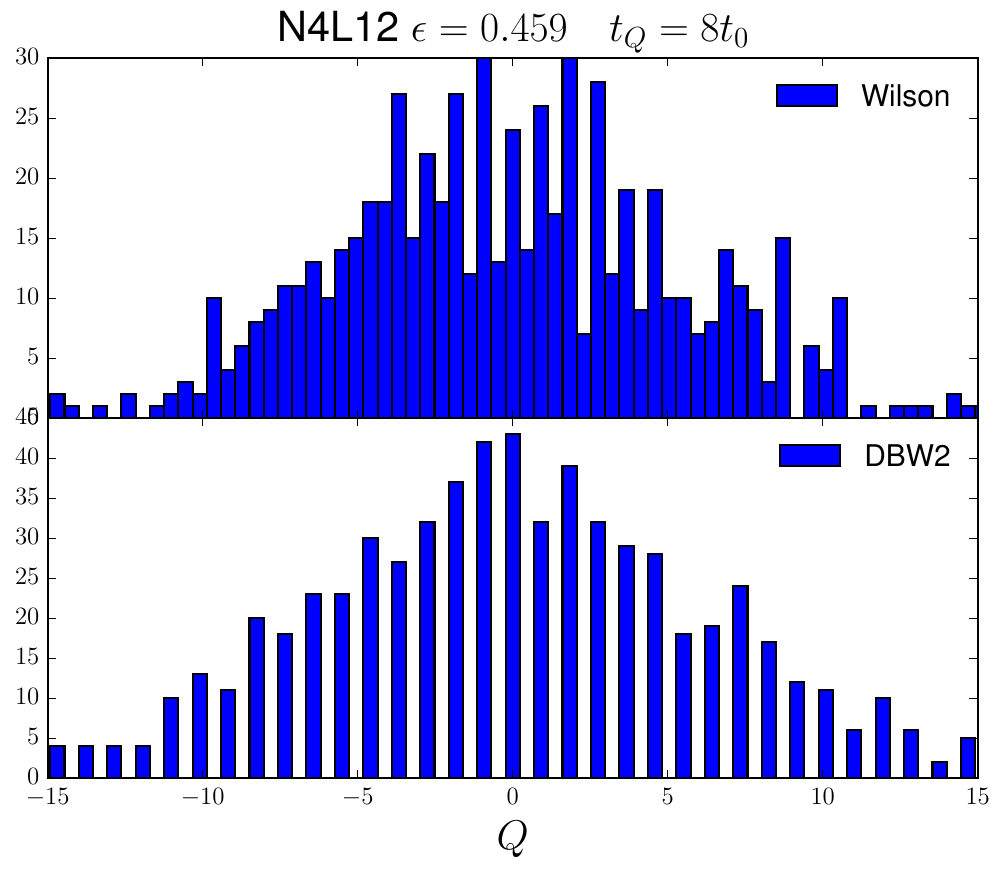}
\caption{Histograms of the topological charge for the Wilson flow (top) and the
  DBW2 flow (bottom) evaluated at flow time $t_Q=2t_0$ (left), $t_Q=4t_0$ (middle)
  and $t_Q=8t_0$ (right) on the \texttt{N4L12} ensemble.}
\label{fig:topohistN4L12}
\end{figure*}

Figure~\ref{fig:topohistN4L12} shows histograms of the topological charge
evaluated at different flow times for the Wilson and the DBW2 flow on the
\texttt{N4L12} ensemble. The bin width is chosen to account for the quantised
units of $Q$ being different from one. In the left panel the histogram of $Q$ is
shown at $t_Q=2t_0$. As can be inferred from Fig.~\ref{fig:compQN4L12} at this
time the Wilson flow has not settled into any asymptotic value yet, leading to
all bins in the central region being populated. The same figure indicates that
for the DBW2 flow, the asymptotic sectors have been reached, i.e. on this
ensemble no crossings between topological sectors occur for later flow times
than $t_Q=2t_0$. However, several configurations still slowly flow to those
discrete values, leading to some contamination between the main bins. For the
DBW2 flow this rapidly improves as the flow time is increased to $t_Q=4t_0$
(middle) and $t_Q=8t_0$ (right). Contrary to this, even for this large flow time
the Wilson flow displays results in the intermediate bins.

\begin{figure*}
\centering
\includegraphics[width=.24\textwidth]{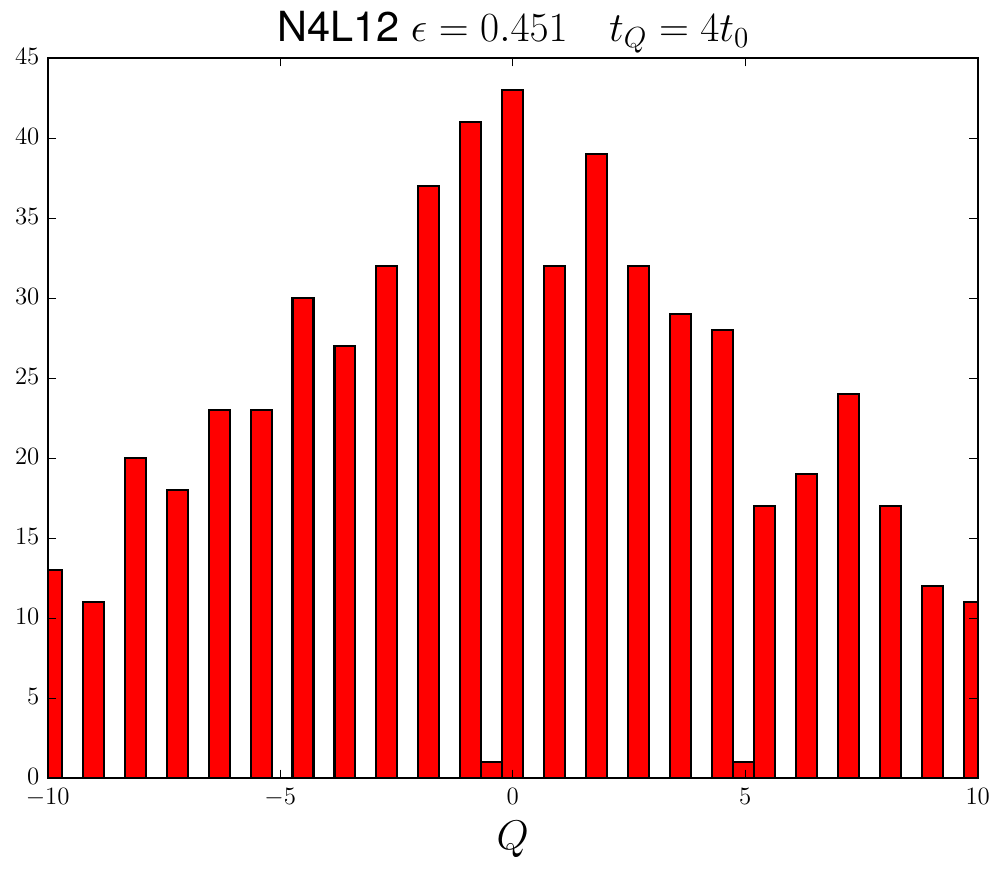}
\includegraphics[width=.24\textwidth]{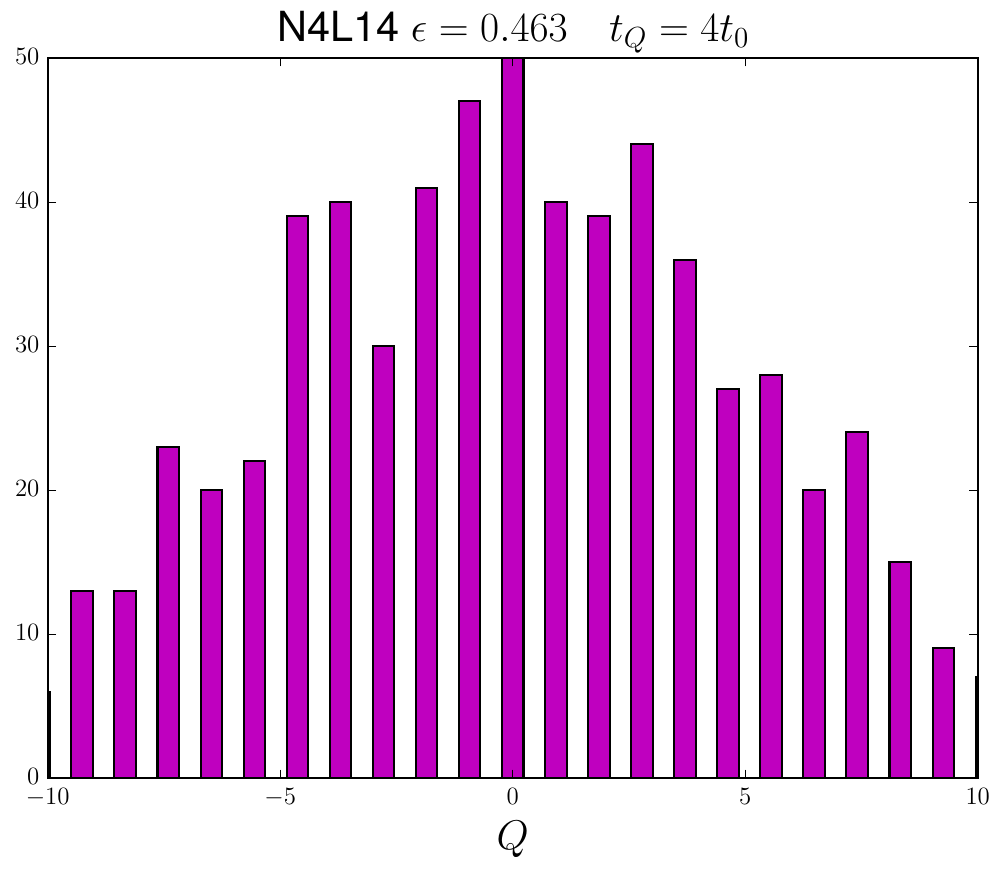}
\includegraphics[width=.24\textwidth]{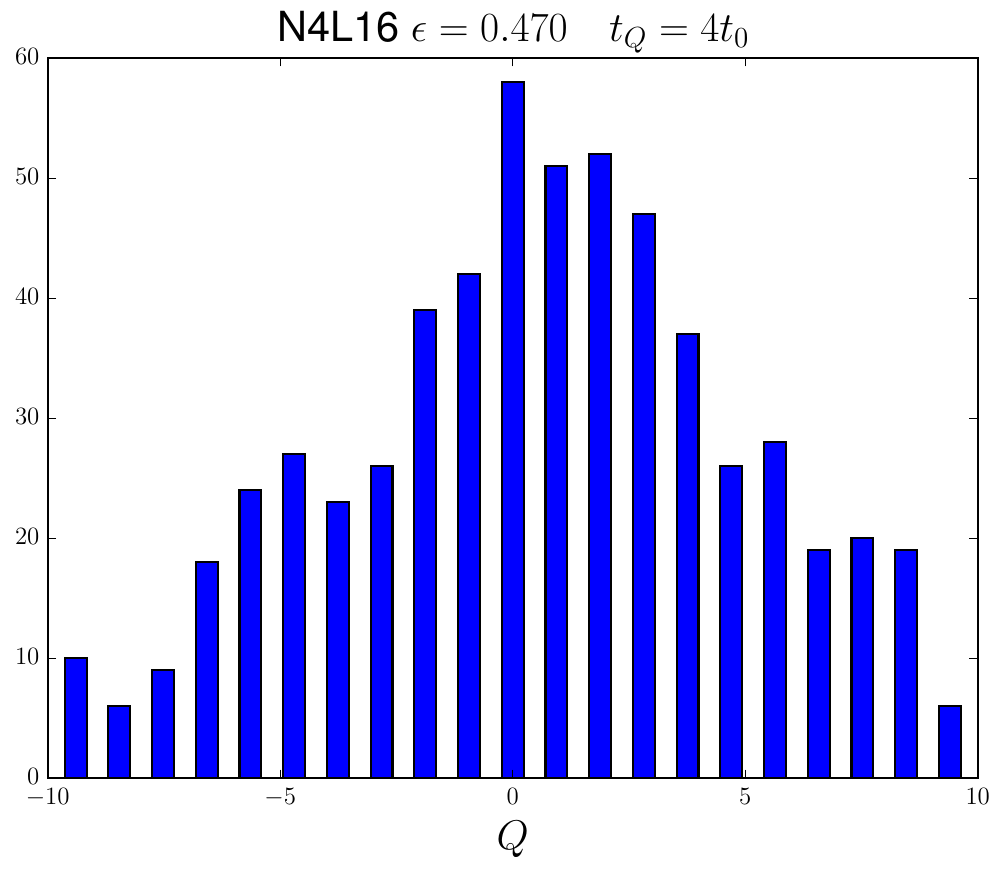}
\includegraphics[width=.24\textwidth]{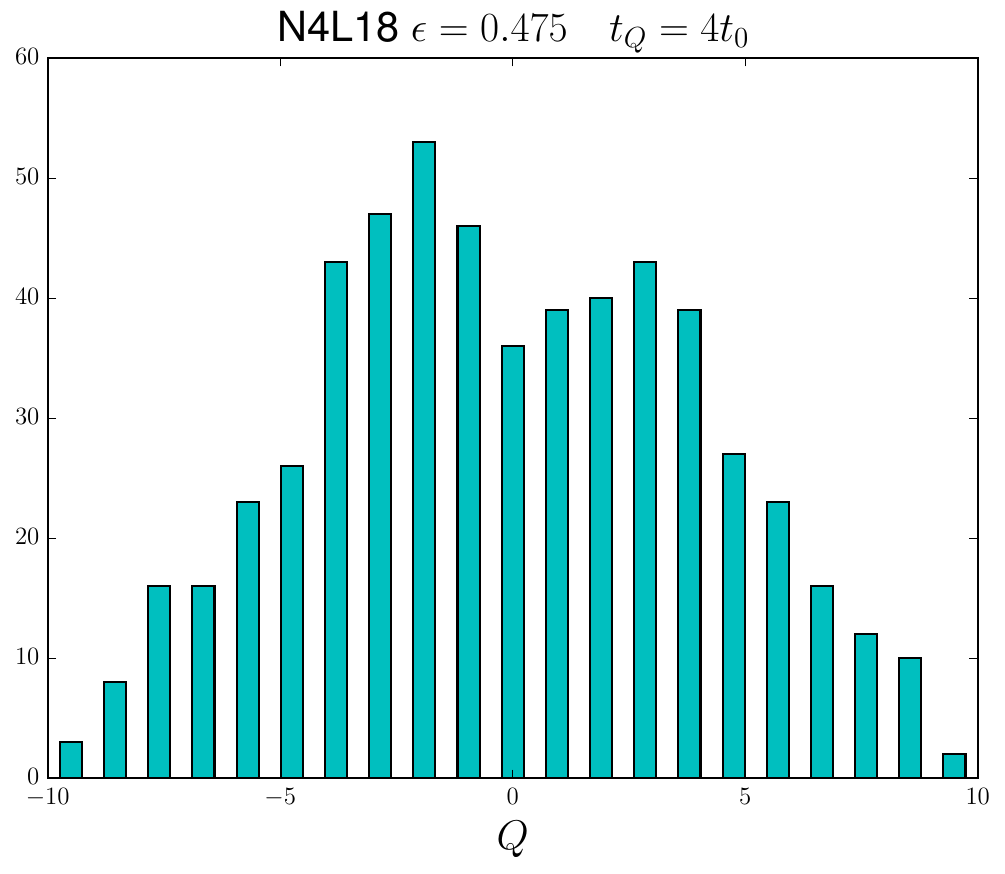}\\
\includegraphics[width=.24\textwidth]{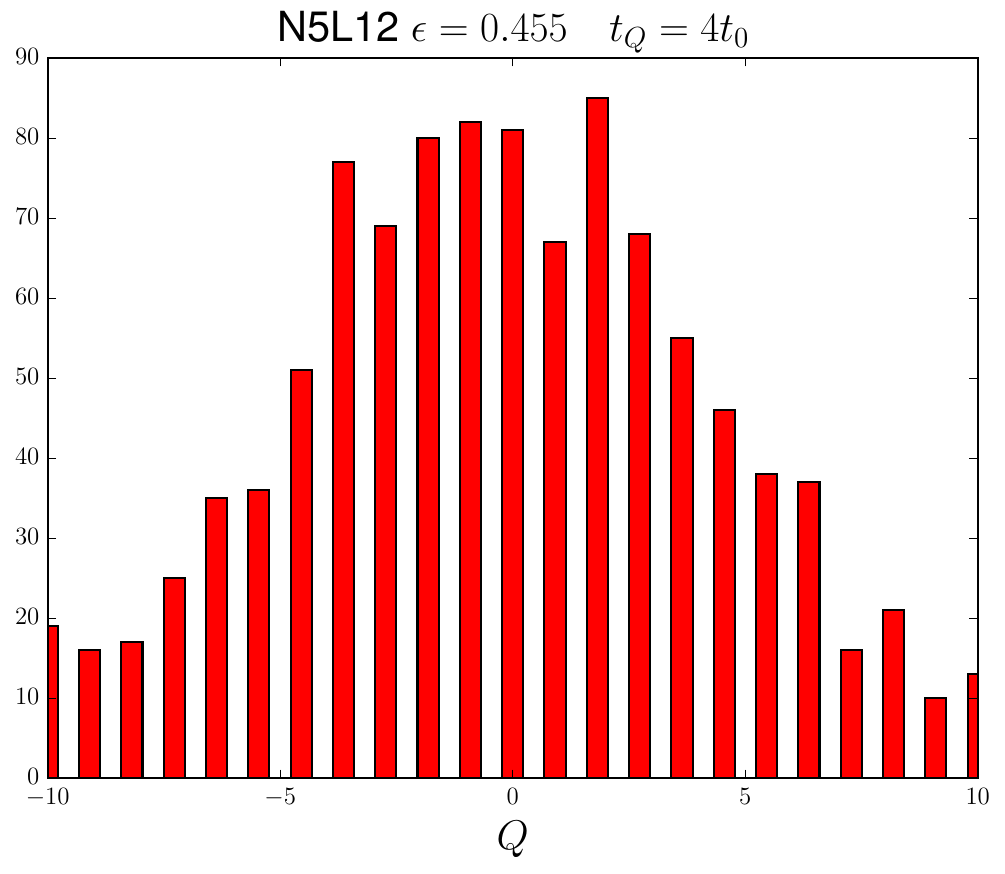}
\includegraphics[width=.24\textwidth]{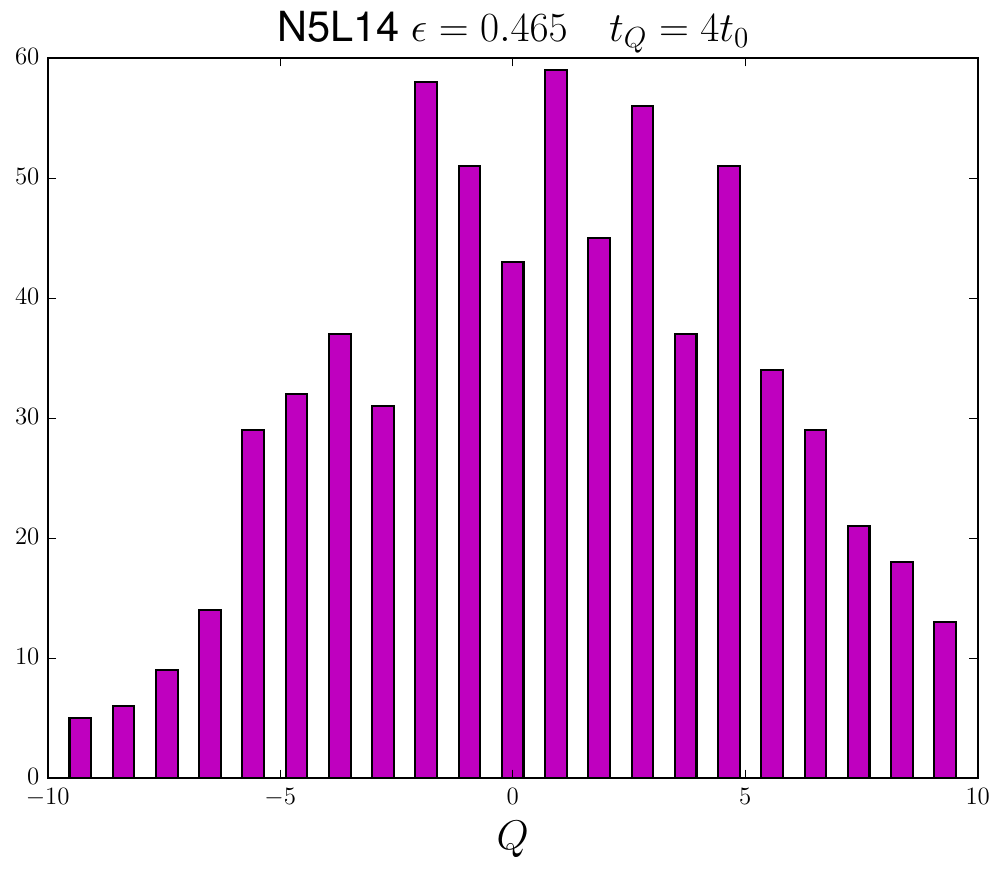}
\includegraphics[width=.24\textwidth]{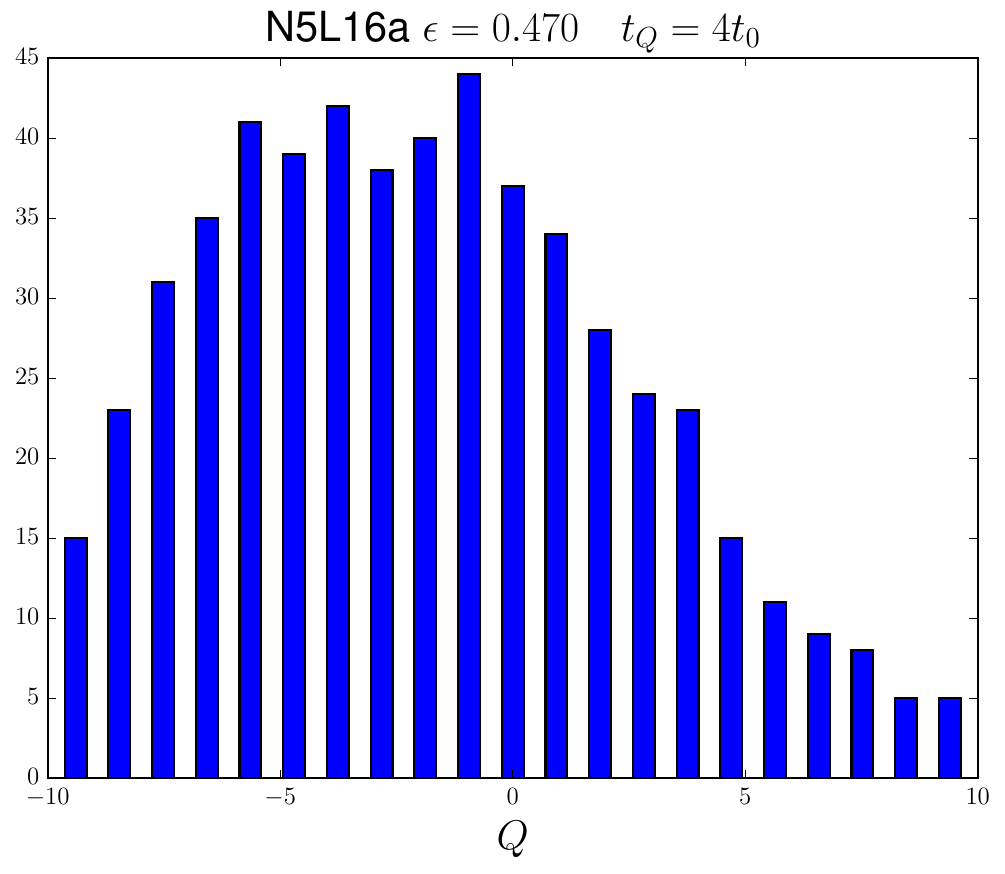}
\includegraphics[width=.24\textwidth]{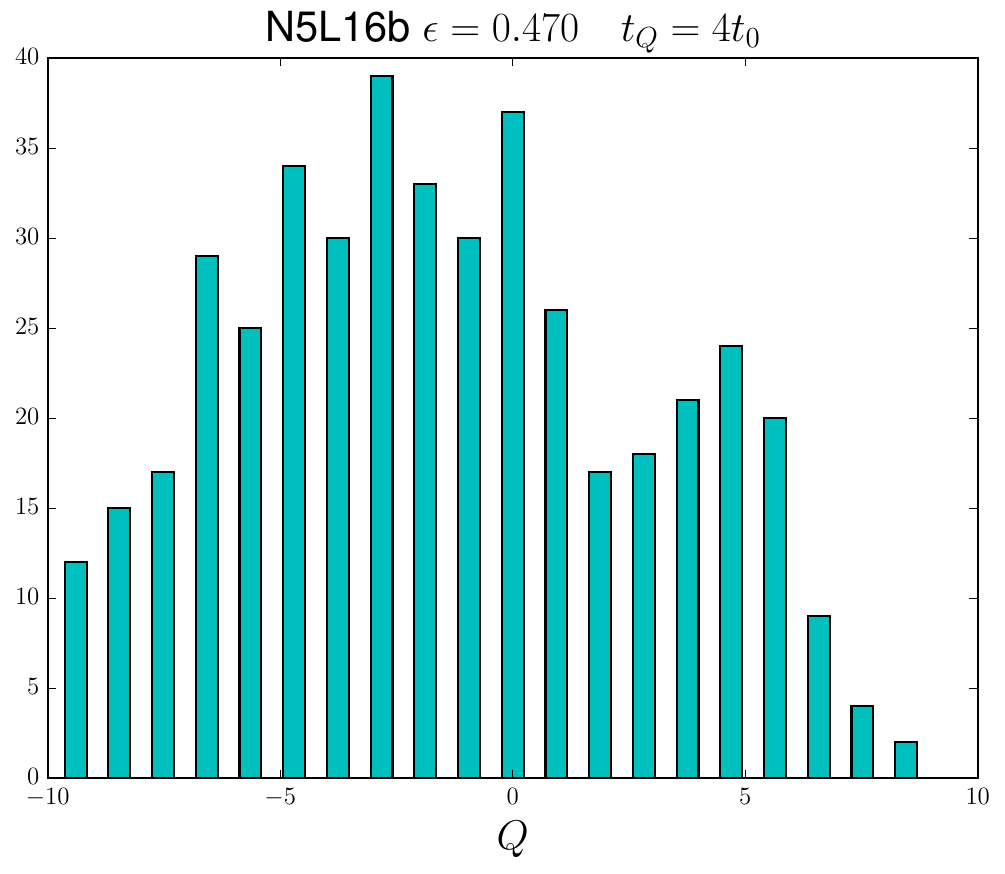}\\
\includegraphics[width=.24\textwidth]{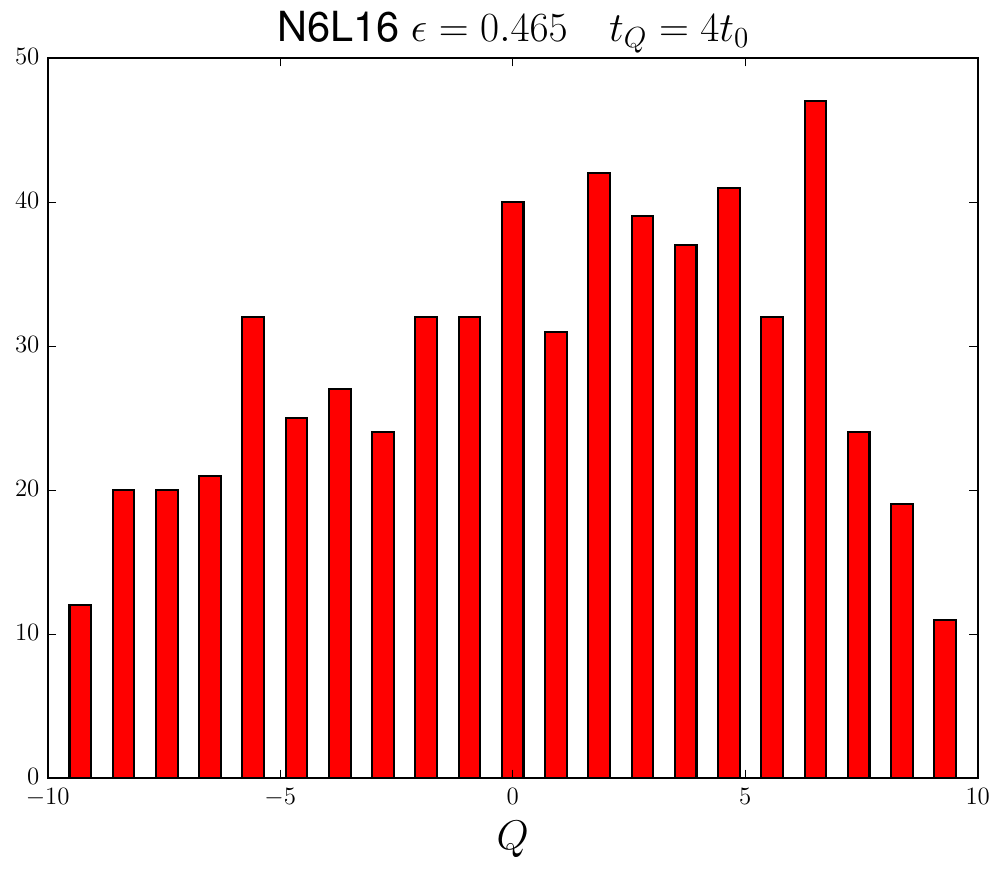}
\caption{Histograms of the topological charge defined via the DBW2 flow at $t =
  4t_0$ for all ensembles considered in this study. From top to bottom $N_C$ is
  increased from 4 to 6 and within each row the lattice spacing is reduced from
  left to right.}
\label{fig:topohistDBW2_all}
\end{figure*}

In Figure~\ref{fig:topohistDBW2_all} we present the histograms of the
topological charge for all ensembles considered in this work. In each case the
value of $\epsilon$ was tuned to ensure that the discrete bins are correctly
captured. We find that $Q$ does take discrete near-integer values in all cases
and hence do not observe any fractional topological charges. We further find
that the unit of discretisation approaches 1 as the lattice spacing is reduced.

\subsection{Dependence of gauge averages on flow smoothening}

We now turn to assessing the impact of the effects discussed in the previous
section on physical observables such as the topological susceptibility and
related quantities. Similar studies of systematic effects associated with the
smoothening radius -- particularly in the context of comparisons between cooling
and gradient flow -- can be found in
Refs.~\cite{Bonati:2014tqa,Bonanno:2023ple,Alexandrou:2015yba,Alexandrou:2017hqw}. In
the same spirit, our goal is to investigate whether variations in the flow time
$t_Q$ or the choice of kernel action introduce significant artefacts on physical
observables.

On our ensembles, i.e. at finite lattice spacing and volume, a direct
computation of the topological susceptibility in the sense of Eq.~\eqref{chi},
which requires the thermodynamic limit, is not possible. Nevertheless, we
compute the quantity $\langle Q^2 \rangle/V$, which we refer to as the
finite-volume topological susceptibility $\tilde\chi$, and compare the results
obtained using the two flows. For each case, we consider several values of the
flow time $t_Q \in \{t_0,2t_0,4t_0,8t_0\} $ to assess the impact on the
extracted values of $\tilde\chi$. We also determine the ratio of
susceptibilities computed with different flows, defining
\begin{equation}
R = \frac{\tilde\chi^{\text{DBW2}}}{\tilde\chi^{\text{Wilson}}}\,,
\end{equation}
which we expect to be less sensitive to slight mistunings of volumes and masses.

\begin{figure}
    \centering
    \includegraphics[width=\linewidth]{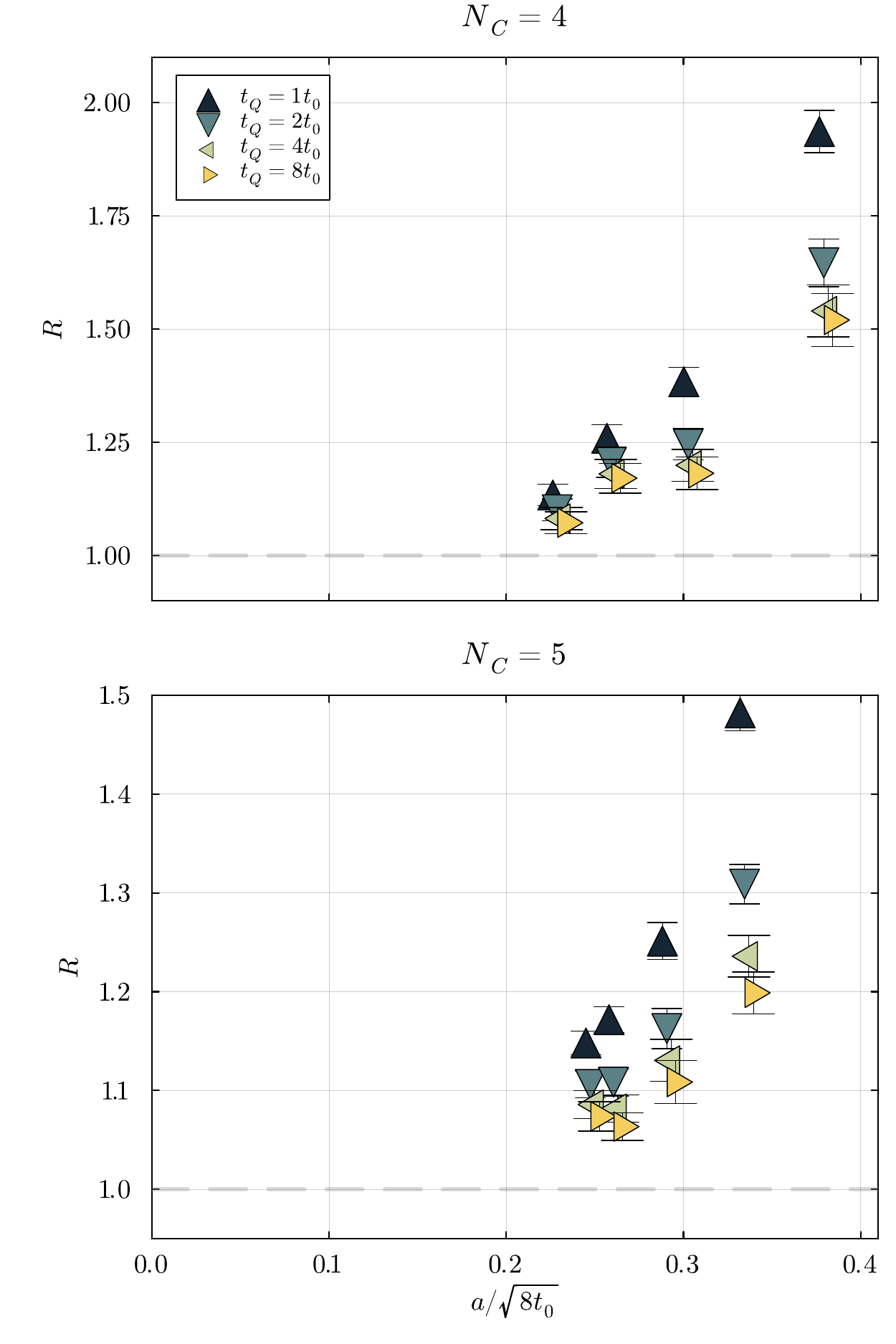}
    \caption{Ratios of susceptibilities computed with Wilson and DBW2 as a
      function of the lattice spacings for $N_C=4$ (upper panel) and $N_C=5$
      (lower panel). As in the caption, the colour code reflects a different
      choice for $t_Q$. Point markers are slightly shifted in the $x$-axis to
      improve visibility.}
    \label{fig:R_vs_a2}
\end{figure}
\begin{figure}
     \centering
     \includegraphics[width=\linewidth]{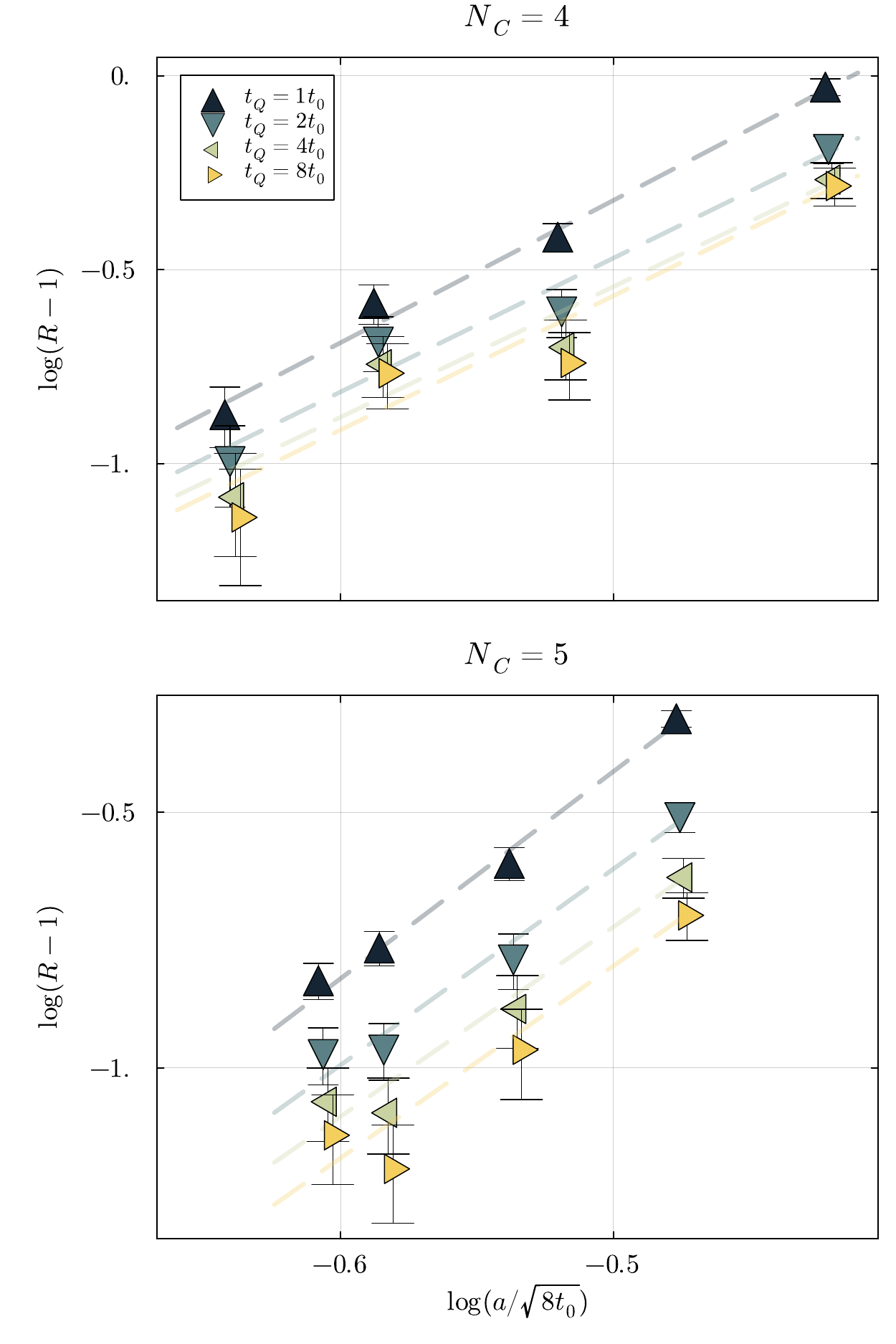}
     \caption{Same as in Fig.~\ref{fig:R_vs_a2}, but in log-log scales. Dashed
       lines are linear fits to the rescaled points, in the caption, the
       exponents of $a$ represent the fitted slopes of the lines.}
     \label{fig:R_vs_a}
 \end{figure}

In Fig.~\ref{fig:R_vs_a2} we plot this ratio as a function of the lattice
spacing in units of $\sqrt{8t_0}$ for our ensembles with $N_C=4$ and $N_C=5$.
We observe that the ratio approaches unity as the lattice spacing is reduced, in
line with our expectations. However, given the coarseness of our lattice
spacings, a linear extrapolation in $a$ is not sufficient, suggesting that
higher powers of $a$ are required to describe the scaling behaviour. To
investigate this, we present the same data in a log-log plot in
Fig.~\ref{fig:R_vs_a}, allowing us to extract the dominant power-law behaviour
in the parameter range covered in this work. A linear fit in this representation
yields an effective scaling exponent between 3 and 4 for both choices of $N_C$,
consistent across all $t_Q$.

\section{Conclusions and Outlook \label{sec:conc}}
We generated ensembles with different lattice spacings but approximately tuned
spatial extents for $SU(4)$, $SU(5)$ and $SU(6)$, with one fermion in the
two-index anti-symmetric representation. On these ensembles we studied the
behaviour of different gradient-flows in the context of the topological
charge. In order to investigate the presence of fractional topological charges
we advocate the use of an over-improved action for the gradient flow as it
quickly and unambiguously settles into discrete topological sectors.  With such
a prescription, the flow-time bounds defined in Eq.~\eqref{eq:bounds} can be
satisfied and a topological charge (and functions of it, such as the topological
susceptibly) can easily be defined. Whilst we do not observe any fractional
topological charges, at finite lattice spacing we find the unit of
discretisation to be slightly below one. We expect the ratio of topological
susceptibilities defined though the Wilson and the over-improved DBW2 flow to
differ from unity by discretisation effects. Our data confirm this and that the
ratio approaches one as the lattice spacing is reduced. We expect these findings
to be independent of the choice of fermion representation used in this work.

Having gained experience in the largely unexplored parameter space for these
theories and defined and determined the topological charge on the ensembles at
hand, we are continuing to generate a suite of ensembles for $N_C > 3$ which
allows the computation of the mesonic spectrum as well as its extrapolation to
the massless limit. This will provide further non-perturbative insight into the
conjectured connection of these theories to Super-Yang-Mills. One crucial step
to achieving this is to define a cost-efficient measurement strategy for the
computation of disconnected diagrams in higher representations, which is ongoing
work. Furthermore, the large-$N_C$ limit of $\mathcal{N}=1$ SUSY Yang-Mills has
been explored for very large $N_C$ using twisted volume reduction
techniques~\cite{Butti:2022sgy,Bonanno:2024bqg,Bonanno:2024bwr,Bonanno:2024onr}. In
the future we plan to explore comparisons between such an approach and the
method presented in this work.

\section*{Acknowledgments}
We are thankful to Fernando Panadero for providing instructions and template
code to use \texttt{LatticeGPU}. We are grateful to Claudio Bonanno, for many
useful discussions and general feedback on the topic discussed here and to Jorge
Dasilva for discussion at the early stage of this work. In addition, we would
like to thank Timo Eichhorn and Antonio Gonz{\'a}lez-Arroyo for their helpful
discussion at Lattice2024 and Tobias Rindlisbacher at the Nordic Lattice meeting
in Lund.

This project has received funding from the European Union’s Horizon 2020
research and innovation program under the Marie Sk{\l}odowska-Curie grant
agreement number 813942. The work of P.B. is supported by the Carlsberg
Foundation, grant CF22-0922.  We gratefully acknowledge the computing time
provided through our allocations. Parameter tuning was performed on the
Discoverer supercomputer under grant EHPC-REG-2023R01-102. The configurations
were generated on LUMI supercomputer under allocation EHPC-EXT-2024E01-038, with
additional support from the DeiC grant (case number 4265-00016A). Additional
support was provided by the UCloud DeiC Interactive HPC system managed by the
eScience Center at the University of Southern Denmark.

\bibliography{paper.bib}

\end{document}